\documentclass[aip,reprint,apl]{revtex4-1}

\usepackage{amsmath,amssymb}
\usepackage{bm}
\usepackage[dvipdfm]{graphicx}
\usepackage[
  dvipdfm,
  bookmarks=true,
  bookmarksnumbered=true,
  bookmarkstype=toc,
  pdftitle={},
  pdfsubject={},
  pdfauthor={Motoaki BAMBA}
]{hyperref}

\newcommand{\ket}[1]{|#1\rangle}
\newcommand{\braket}[1]{\langle #1 \rangle}

\def\ee{\mathrm{e}}
\def\ii{\mathrm{i}}

\def\Hc{\mathrm{H.c.}}

\def\wp{\omega_{\text{p}}}
\def\DE{\delta E}
\def\Udir{U}
\def\Ucross{U_{\text{cross}}}

\def\oH{\hat{H}}
\def\oa{\hat{a}}
\def\oad{\hat{a}^{\dagger}}

\begin{document}


\title{Counter-polarized single-photon generation
from the auxiliary cavity \\ of a weakly nonlinear photonic molecule}

\author{Motoaki Bamba}
\altaffiliation{E-mail: motoaki.bamba@univ-paris-diderot.fr}
\affiliation{Laboratoire Mat\'eriaux et Ph\'enom\`enes Quantiques,
Universit\'e Paris Diderot-Paris 7 et CNRS, \\ B\^atiment Condorcet, 10 rue
Alice Domon et L\'eonie Duquet, 75205 Paris Cedex 13, France}
\author{Cristiano Ciuti}
\altaffiliation{E-mail: cristiano.ciuti@univ-paris-diderot.fr}
\affiliation{Laboratoire Mat\'eriaux et Ph\'enom\`enes Quantiques,
Universit\'e Paris Diderot-Paris 7 et CNRS, \\ B\^atiment Condorcet, 10 rue
Alice Domon et L\'eonie Duquet, 75205 Paris Cedex 13, France}

\date{\today}

\begin{abstract}
We propose a scheme for the resonant generation of counter-polarized single photons
in double asymmetric cavities with a small Kerr optical nonlinearity (as that created by a semiconductor quantum well) compared to the mode broadening.
Due to the interplay between spatial intercavity tunneling and polarization coupling, by weakly exciting with circularly polarized light one of the cavities,
we predict strong antibunching of counter-polarized light emission from the non-pumped auxiliary cavity.
This scheme due to quantum interference is robust against surface scattering of pumping light,
which can be suppressed both by spatial and polarization filters. 
\end{abstract}


\maketitle
Single photons play an essential role in quantum information technologies
and their generation is a fascinating subject
in the fields of quantum optics and condensed matter physics.
As deterministic (on-demand) sources of single photons, 
single semiconductor quantum dots
embedded in optical microcavities have been investigated with
impressive results  in the case of non-resonant  excitation.
\cite{Michler2000S,Moreau2001PRL,Vuckovic2003APL,Press2007PRL,Shields2007NP,Claudon2010NP}
However, if one wishes to build an  array of single-photon emitters with intentional 
pattern on  the same wafer, the completely random position of self-organized quantum dots is 
by definition a kind of limitation. In the case of non-resonant generation, the repetition rate of the single-photon
source is limited by the relatively slow relaxation time of the injected carriers in the
semiconductor device. 
For higher repetition rates and freedom in array design, the resonant photon blockade in photonic pillars
including a Kerr medium has been proposed, where the nonlinear medium is a simple semiconductor quantum well,\cite{verger06}
which can be patterned with great flexibility. 
Unfortunately, the non-trivial difficulty of such approach is that the strength of the Kerr nonlinearity must be much larger than
the mode broadening, thus requiring in practice the use of ultrasmall photonic resonators with very high quality factors.

In a recent letter,\cite{Liew2010PRL} it has been proved that by using two coupled pillars (a double cavity system
or photonic `molecule') it is possible to get very pure single-photon emission with a nonlinearity surprisingly small
with respect to the losses.
As analytically shown 
in Ref.~\onlinecite{Bamba2011Antibunching},
the antibunching originates from the destructive quantum interference
between excitation and tunneling paths of photons.
However, a practical hurdle still remains: in the proposed scheme, \cite{Liew2010PRL,Bamba2011Antibunching}
single photons are emitted from the excited cavity with the same polarization of the pump.
Hence, spurious effects as pump surface scattering could mask such an effect. 
In the present paper, we propose another scheme based on the generalization 
of the photonic molecule approach by taking advantage of the polarization degree of freedom
in asymmetric cavities having a frequency splitting between modes with orthogonal linear polarizations.
We show how to get counter-polarized single-photons from the non-pumped auxiliary cavity, thus providing a way
to get rid of the pump spurious scattering by both spatial and polarization filtering. 

\begin{figure}[tbp]
\includegraphics[width=\linewidth]{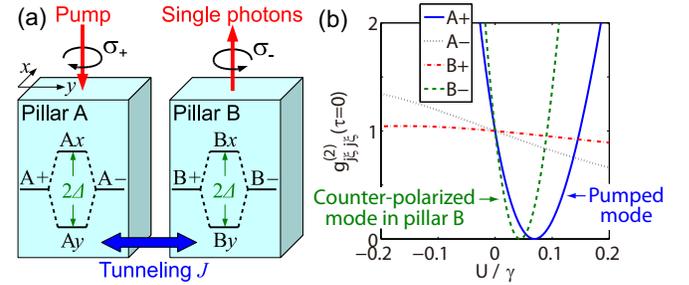}
\caption{(a) Sketch of the system consisting of two coupled micropillars: due to the shape anisotropy, the photon eigenmodes
have orthogonal linear polarizations.
The mode with $\xi$ polarization in pillar $j$
is denoted as $j\xi$, and the energy levels for the single photon states in each pillar are depicted.
By illuminating circularly polarized light on pillar A,
counter-polarized single photons
are emitted from pillar B even with a small nonlinearity.
(b) The equal-time second-order correlation functions
$\{g^{(2)}_{j\xi j\xi}(\tau=0)\}$ are plotted
as functions of nonlinearity $U$ normalized to broadening $\gamma$.
The tunneling strength is $J = 5\gamma$,
the polarization splitting is $\varDelta = 2.5\gamma$,
and the pump frequency is tuned as $\DE = E - \hbar\wp = 0.2772\gamma$.
In addition to the antibunching of the pumped mode $A+$,
nearly perfect antibunching is obtained in mode $B-$
with the relatively small nonlinearity $U = 0.0438\gamma$.}
\label{fig:1}
\end{figure}
As a realistic system,
we consider two spatially separated semiconductor micropillars with asymmetric shape,\cite{bajoni08} coupled with a photonic tunneling strength $J$.
Each pillar has two different linearly polarized photonic modes
($x$ and $y$ directions),
energetically split by $2\varDelta$
due to the shape anisotropy.
Fig.~\ref{fig:1}(a) shows a sketch of considered system.
The Hamiltonian is represented as
\begin{align}
\oH
& = \sum_{j=\{A,B\}}\left[ (E+\varDelta)\oad_{jx}\oa_{jx}
    + (E-\varDelta)\oad_{jy}\oa_{jy} \right]
\nonumber \\ & \quad
+ \sum_{\xi=\{x,y\}}J(\oad_{A\xi}\oa_{B\xi} + \Hc)
+ (F \ee^{-\ii\wp t} \oad_{A+} + \Hc)
\nonumber \\ & \quad
+ \sum_{j=\{A,B\},\xi=\{+,-\}}
    \Udir\oad_{j\xi}\oad_{j\xi}\oa_{j\xi}\oa_{j\xi}.
\label{eq:Hamiltonian} 
\end{align}
Here, $\oa_{j\xi}$ is the annihilation operator of a photon
with polarization $\xi$ in pillar $j$. The relation between circularly and linearly polarized modes 
is given by the standard operator expression  $\oa_{j\pm} = (\oa_{jx} \pm \ii\oa_{jy}) / \sqrt{2}$.
We consider the configuration where pillar A is pumped with $\sigma_+$-circular polarization, being   $\wp$ and $F$ the pump frequency and amplitude
respectively.
The pumping strength is moderate
to guarantee the average number of photons
in the system not exceeding unity.
If the average number is increased, antibunching is worsened,
because the quantum interference in the present scheme is valid
if three-photon subspace can be neglected, as in the previously considered scenario.
\cite{Liew2010PRL,Bamba2011Antibunching}
The nonlinearity is represented by the last term
in Eq.~\eqref{eq:Hamiltonian}
conserving the total spin of two photons. 
The effective nonlinearity can be mediated by the presence of a quantum well excitonic resonance,\cite{Ciuti1998PRB}
but this effective Kerr term is quite general for systems with a third-order nonlinearity.
The cross-polarized term such as $\Ucross\oad_{j+}\oad_{j-}\oa_{j-}\oa_{j+}$
is not considered in the present paper,
because $\Ucross$ is usually much smaller than $\Udir$,\cite{Amo2010PRB} but it could be
added without qualitative changes (not shown).
By using the theoretical method detailed in Ref.~\onlinecite{verger06},
we have numerically calculated second-order correlation
functions $g^{(2)}_{j\xi j'\xi'}(\tau) = \braket{\oad_{j\xi}\oad_{j'\xi'}(\tau)\oa_{j'\xi'}(\tau)\oa_{j\xi}}/\braket{\oad_{j\xi}\oa_{j\xi}}\braket{\oad_{j'\xi'}\oa_{j'\xi'}}$
at the steady state under continuous pumping
and a dissipation of photons with a rate $\gamma/\hbar$
in each mode.

\begin{figure}[tbp]
\includegraphics[width=\linewidth]{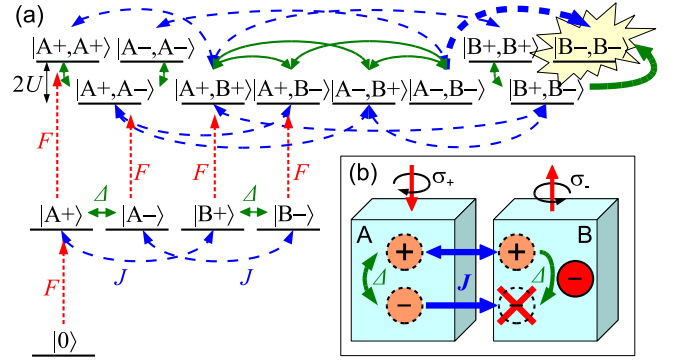}
\caption{(a) Sketch of all the transition paths between zero-photon $\ket{0}$,
one-photon $\ket{j\xi}$, and two-photon states $\ket{j\xi,j'\xi'}$.
The antibunching of mode $B-$ is due to the destructive quantum
interference between the two paths from $\ket{A-,B-}$ and $\ket{B+,B-}$
to $\ket{B-,B-}$.
(b) Pictorial representation. 
If there is already a ``$-$'' photon in pillar B,
another ``$-$'' photon cannot exists in the same pillar
because of the interference
between the $J$-assisted (spatial tunneling) path from ``$-$'' photon in pillar A
and the $\varDelta$-assisted (polarization coupling) path from ``$+$'' photon in pillar B. This quantum interference
occurs for an optimal value of the nonlinearity and laser detuning.}
\label{fig:2}
\end{figure}
In Fig.~\ref{fig:1}(b), we plot equal-time correlations
$\{g^{(2)}_{j\xi j\xi}(\tau=0)\}$ as a function  of nonlinearity
$\Udir$ normalized to $\gamma$.
We consider the tunneling strength $J = 5\gamma$,
polarization splitting $\varDelta = 2.5\gamma$,
and pump detuning $\DE = E - \hbar\wp = 0.2772\gamma$.
In addition to antibunching at the pumped mode $A+$
due to the previously proposed scheme \cite{Liew2010PRL,Bamba2011Antibunching}
strong antibunching is achieved
for mode $B-$ with the (small) optimal nonlinearity $U = 0.0438\gamma$.
In the weak pumping limit, $g^{(2)}_{B-B-}(\tau=0)$ is reduced to zero.
The underlying  destructive quantum interference is different
from the previous one in Refs.~\onlinecite{Liew2010PRL} and \onlinecite{Bamba2011Antibunching}.
By deriving the equations of motions for the amplitudes of the possible Fock states
for the zero-, one- and two-photon states (generalizing the method in Ref.~\onlinecite{Bamba2011Antibunching})
we have derived the optimal conditions of the counter-polarized antibunching
and found the interference paths
leading to the antibunching in mode $B-$.
In Fig.~\ref{fig:2}(a),
the zero-, one-, and two-photon state manifolds are depicted 
and labeled as
$\ket{0}$, $\ket{j\xi}$, and $\ket{j\xi,j'\xi'}$, respectively.
The paths responsible to antibunching in mode $\ket{B-,B-}$
are the $J$-assisted (spatial tunneling) path from $\ket{A-,B-}$
and the $\varDelta$-assisted (polarization coupling) one from $\ket{B+,B-}$:
destructive interference occurs for an optimal value of the nonlinearity $U$, which is much smaller
than the inverse of the cavity lifetime.
Fig.~\ref{fig:2}(b) shows a pictorial interpretation.
In presence of ``$-$''-photon in cavity B,
another photon cannot enter the same pillar due to the quantum interference.
The presence of a cross-polarized nonlinear term $\Ucross$
would not change this picture (not shown) and it does not create  a new path to $\ket{B-,B-}$.
Furthermore, while identical pillars are supposed in the reported calculation,
we have numerically checked that
nearly perfect antibunching in mode $B-$ can be obtained
even under a deviation of the order of $\gamma$
on eigen frequencies $E$
and splitting $\varDelta$ between the two pillars,
and also tunneling strength $J$ between two polarizations,
while the pumping frequency $\wp$ should be properly tuned.

\begin{figure}[tbp]
\includegraphics[width=\linewidth]{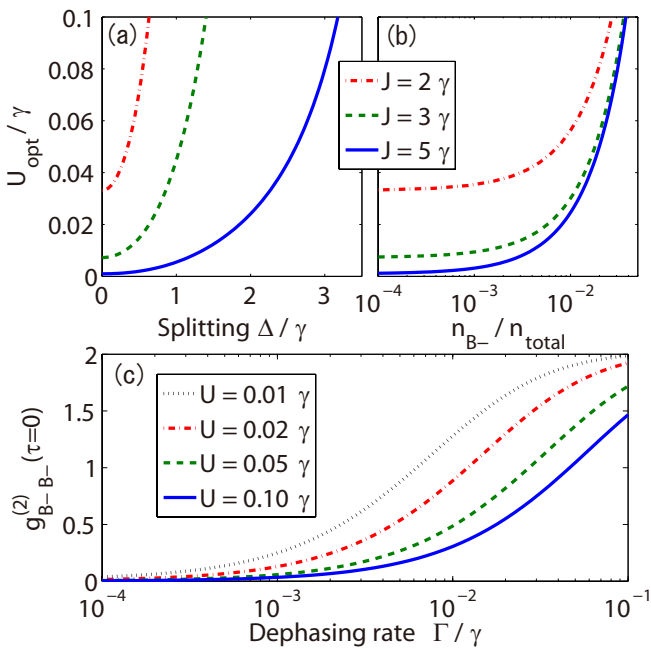}
\caption{(a) Optimal nonlinearities $U_{\text{opt}}$ are plotted
as a function  of $\varDelta/\gamma$
for tunneling strengths $J/\gamma = 2$, 3, and 5.
(b) Under the optimal conditions,
the ratios between the average number $n_{B-}$ of photons in mode $B-$
and the number $n_{\text{total}}$ in the total system
are plotted for corresponding $U_{\text{opt}}/\gamma$.
(c) The obtainable $g^{(2)}_{B-B-}(\tau=0)$
are plotted versus the dephasing rate $\varGamma$
normalized to $\gamma$.
The tunneling strength is $J = 5\gamma$,
 energy splitting $\varDelta$ and pumping frequency
are chosen to give the nearly perfect antibunching
for each nonlinearity $U$ in the absence of pure dephasing.}
\label{fig:3}
\end{figure}

From the equations of motions of up to the two-photon Fock states,
we have calculated the optimal nonlinearity
$U_{\text{opt}}$ for the perfect antibunching
under the weak pumping limit.
Fig.~\ref{fig:3}(a) shows $U_{\text{opt}}$
as a function of $\varDelta/\gamma$,
and Fig.~\ref{fig:3}(b) represents the ratio between
the average photon number $n_{B-} = \braket{\oad_{B-}\oa_{B-}}$
in mode $B-$ and the total number
$n_{\text{total}} = \sum_{j,\xi} \braket{\oad_{j\xi}\oa_{j\xi}}$
for the corresponding $U_{\text{opt}}$.
The minimum nonlinearity that is required for the antibunching is decreased
by the increase of $J$,
and the required nonlinearity is decreased
together with the splitting $\varDelta$
(or the opposite behavior when $J < \varDelta$).
However, together with the decrease of $U_{\text{opt}}$,
the occupation probability of mode $B-$ is significantly decreased
as seen in Fig.~\ref{fig:3}(b).
For $U_{\text{opt}} < 0.1\gamma$,
the curve in Fig.~\ref{fig:3}(b) is almost saturated at $J = 5\gamma$,
and we can obtain a probability
$n_{B-}/n_{\text{total}} \sim 10^{-2}$,
which corresponds to a generation rate
of the order of 100\;MHz for a cavity lifetime in the picosecond rage.
This rate is higher than that of the quantum dots
\cite{Michler2000S,Moreau2001PRL,Vuckovic2003APL,Press2007PRL,Shields2007NP,Claudon2010NP}
by one order of magnitude.

Finally, we have examined the robustness of the present scheme
against dephasing of photons.
Since quantum interferences are responsible in the present
and previous schemes,\cite{Liew2010PRL,Bamba2011Antibunching}
pure dephasing can decrease the quality of the antibunching.
By using the standard pure dephasing model due to quadratic coupling
with a reservoir,\cite{Walls1985PRD}
we consider dephasing with a rate $\varGamma/\hbar$
affecting linearly polarized modes of each pillar
(the results shown below are not significantly modified
even if the dephasing is supposed to affect the circularly polarized modes).
Fig.~\ref{fig:3}(c) shows $g^{(2)}_{B-B-}(\tau=0)$
as a function of $\varGamma/\gamma$.
The tunneling strength is $J = 5\gamma$,
and the splitting $\varDelta$ is chosen to give perfect antibunching
for each nonlinearity $U$ in the absence of the dephasing.
As clearly shown, the antibunching
can be significantly worsened in presence of pure dephasing for a given
value of the nonlinearity.
However, even if the nonlinearity is quite small,
for example $U = 0.05\gamma$,
antibunching is still observable if $\varGamma = 10^{-2}\gamma$
and becomes very strong if $\varGamma = 10^{-3}\gamma$.
The curves in Fig.~\ref{fig:3}(c) does not strongly depend on
the tunneling strength $J$
if the nonlinearity $U$ is large enough
compared to the corresponding minimum shown in Fig.~\ref{fig:3}(a).
Since the optimal nonlinearity can be weak in the present scheme, one can
consider cavities with relatively small photon lifetime (small quality factor)
in a regime where the pure dephasing time can be thus neglected, a very
promising outlook.

In conclusion,
we have proposed a scheme of single-photon generation
due to a destructive quantum interference effect
in a weakly nonlinear double cavity system, where
each cavity have two linearly polarized, frequency-split modes.
Due to the spatial tunneling between the two cavities
and the coupling between opposite circular polarizations,
we have found that strong antibunching of counter-polarized emission
can be obtained at the non-pumped auxiliary pillar. 
This new effect is of practical implications, because it provides
a direct way to suppress the pump scattering via spatial and polarization filters.
This scheme can also exported to arrays
of nonlinear photonic molecules.

We would like to thank A. Amo, A. Bramati, J. Bloch, I. Carusotto, E. Giacobino, and A. Imamoglu
for stimulating discussions. We acknowledge support from ANR grants SENOQI and QPOL. 
C. Ciuti is member of Institut Universitaire de France (IUF).

\end{document}